\documentstyle[twocolumn,eqsecnum,aps]{revtex}
\newcommand{\beq}{\begin{equation}}
\newcommand{\eeq}{\end{equation}}
\newcommand{\beqa}{\begin{eqnarray}}
\newcommand{\eeqa}{\end{eqnarray}}

\newcommand{\kBT}{\mbox{$k_{\rm B}T$}}

\newcommand{\del}{\partial}
\newcommand{\G}{{\cal G}} 
\renewcommand{\vec}[1]{\mbox{\boldmath$#1$}}
\newcommand{\svec}[1]{\mbox{\small\boldmath$#1$}} 

\begin{document}
%\large
\draft
%\preprint{HEP/123-qed}
\title{Statistical Mechanical Calculation of Anisotropic Step Stiffness
 of a Two-Dimensional Hexagonal Lattice Gas  Model with Next-Nearest-Neighbor
 Interactions: Application to Si(111) Surface
}
\author{Noriko Akutsu$^1$ and Yasuhiro Akutsu$^2$\\
}
\address{
$^1$Faculty of Engineering, Osaka Electro-Communication
 University,
Hatsu-cho, Neyagawa, Osaka 572, Japan\\
$^2$Department of Physics, Graduate School of Science, Osaka University,
Machikaneyama-cho, Toyonaka, Osaka 560, Japan
}
\date{\today}
\maketitle
\begin{abstract}
We study a two-dimensional honeycomb lattice gas model with both
 nearest- and next-nearest-neighbor interactions in a staggered field,
 which describes the surface of stoichiometrically binary crystal.
 We calculate anisotropic step tension, step stiffness, and equilibrium
  island shape, by an extended random walk method.  We apply the
 results to Si(111) 7$\times$7 reconstructed surface and high-temperature
 Si(111) 1$\times$1 surface.  We also calculate inter-step interaction
 coefficient.
\end{abstract}
%
%
%\pacs{Valid PACS appear here.
%{\tt$\backslash$\string pacs\{\}} should always be input,
%even if empty.}

%\maketitle
%

\narrowtext

\section{Introduction}

Recent developments of the experimental techniques such as STM
 (scanning-tunneling
 microscopy)\cite{stm},  LEEM (low-energy electron microscopy)\cite{leem}
 and REM (reflection electron microscopy)\cite{rem} make it possible
 to observe a step on a crystal surface in the wide range of the
 length scales.  However, the connection among quantities measured
 in different scales has not been clarified yet.

In Ref.\cite{akutsu97}, for two-dimensional (2D) square-lattice Ising
 model with both nearest- and next-nearest-neighbor (nn and nnn)
 interactions, we calculated anisotropic interface tension and interface
 stiffness by the imaginary path-weight (IPW) method which is an
  extended Feynman-Vdovichenko's random walk
 method\cite{Vdo,holzer,akutsu90,akutsu90j}.
 In the method, the overhang structure in a step is taken into account,
 which leads to high accuracy of the results in a wide range of temperature.
  
We applied the results to Si(001) surface based on the microscopic
 kink energy obtained by the Swartzentruber
 {\it et al.}\cite{swartzentruber}.
  The Ising result gave a satisfactory explanation for experimentally
 measured step tension $\gamma$, step stiffness $\tilde{\gamma}$
 and equilibrium island shape obtained by Bartel
t {\it et al.}\cite{bartelt94}
 on Si(001) surface.

 In the present paper, we consider the honeycomb lattice Ising system
 in a staggered field, with both nn and nnn interactions, to calculate
 interface tension, interface stiffness, island shape and the coefficient
 of step interaction by the IPW method.  We aim at applying the results
 to  Si(111) 7$\times$7 reconstructed surfaces and high-temperature
 Si(111) 1$\times$1 surface.

\section{Model Hamiltonian}
We consider a honeycomb lattice with $2N$ sites.  We decompose the
 lattice into two triangular sublattices designated by A and B. 
   On the A-sublattice, we define the occupation variable $C_{{\rm
 A}i}$ which takes 1 (present) or 0 (absent) at the site $i$.   Similarly,
 we define  $C_{{\rm B}j}$ for the B-sublattice.  

  The lattice gas Hamiltonian ${\cal H}_{\rm LG}$ is then written
 as
\beqa
	{\cal H}_{\rm LG} = - 4J_1\sum_{<i,j>}[C_{{\rm A}i}C_{{\rm
 B}j}- \frac{1}{2}(C_{{\rm A}i} + C_{{\rm B}j})] \nonumber \\ 
- 4J_{\rm A2}\sum_{<i,j>}[C_{{\rm A}i}C_{{\rm A}j}- \frac{1}{2}(C_{{\rm
 A}i} + C_{{\rm A}j})] \nonumber \\
- 4J_{\rm B2}\sum_{<i,j>}[C_{{\rm B}i}C_{{\rm B}j}- \frac{1}{2}(C_{{\rm
 B}i} + C_{{\rm B}j})] \nonumber \\
 - \epsilon_{{\rm A}}\sum_{i=1}^{N} C_{{\rm A}i} -  \epsilon_{{\rm
 B}}\sum_{i=1}^{N} C_{{\rm B}i} , %\tag{2.1}
\eeqa
where  $4J_1$ is the bond energy between A-atom and B-atom of the
 nn sites,  $4J_{\rm A2}$, and $4J_{\rm B2}$ are the bond energies
 between  nnn atoms.  $\epsilon_{\rm A}$($\epsilon_{\rm B}$) is the
 ``surface chemical potential'' of A-atom (B-atom).  We consider
 the simplest case where $\epsilon_{\rm A}$ and $\epsilon_{\rm B}$
 are given by
\beqa
\left. \begin{array}{ll}
 \epsilon_{\rm A} = &\mu_{\rm A,gas}(P_{\rm A}, P_{\rm B}, T) - \mu_{\rm
 A,surf}(T) ,\\
 \epsilon_{\rm B} = & \mu_{\rm B,gas}(P_{\rm A}, P_{\rm B}, T) -
 \mu_{\rm B,surf}(T) . 
\end{array}
\right\}  %\tag{2.2}
\eeqa
In the above, $\mu_{\rm A,gas}(P_{\rm A}, P_{\rm B}, T)$ is the chemical
 potential of A-atom  in the gas, $P_{\rm A}$ is the partial pressure
 of A-atom in the gas phase (similarly for $\mu_{\rm B,solid}(T)$,
 $\mu_{\rm B,gas}(P_{\rm A}, P_{\rm B}, T)$ and $P_{\rm B}$);  $\mu_{\rm
 A,surf}(T) $ and $\mu_{\rm B,surf}(T) $  are expressed as
\beqa
  \mu_{\rm A,surf} = &\Delta E( T) + \mu_{\rm solid}(T) , \nonumber
 \\
  \mu_{\rm B,surf} = &-\Delta E( T) + \mu_{\rm solid}(T) ,
%\tag{2.3}
\eeqa
where  $\Delta E( T)$ has been introduced as the difference from
 the chemical potential of atoms in the bulk solid $\mu_{\rm solid}(T)$.

Let us consider the bulk phase-coexistence state of the stoichiometrically
 binary system.   Total chemical potential of the system has to be
 unchanged under removal  of one pair of AB atoms from crystal into
 vapor and vice-versa.
  Hence,  as the coexistence condition, we have
\beqa
 \mu_{\rm A,gas}(P_{\rm A}, P_{\rm B}, T) + \mu_{\rm B,gas}(P_{\rm
 A}, P_{\rm B}, T) \nonumber \\
 = 2  \mu_{\rm solid}(T). %\tag{2.4}
\eeqa
Combining  (2.2)-- (2.4), we obtain
\beq
\epsilon_{\rm B} = -\epsilon_{\rm A}. 
%\tag{2.5}
\eeq
This condition also means that, in the lattice gas Hamiltonian (2.1),
 the total energy of the all-occupied state is the same at that of
 the all-empty state.

Let us introduce the Ising spin variables \{$\sigma_{{\rm A} i} $\}
 and \{$\sigma_{{\rm B}i} $\}  as
\beq
\sigma_{{\rm A} i} = 2\left( C_{{\rm A} i} - \frac{1}{2} \right),\quad
 \sigma_{{\rm B} i} = 2\left( \frac{1}{2} - C_{{\rm B}i}\right).
%\tag{2.6}
\eeq
Substituting (2.6) into the Hamiltonian (2.1) together with (2.5),
 we have the Ising AF Hamiltonian ${\cal H}$:
\beqa
{\cal H} &=& {\cal H_{\rm LG}} + Nz_6J_1 + Nz_3(J_{\rm A2}+J_{\rm
 B2})/2,\nonumber \\
 &=&  J_1\sum_{<i,j>}\sigma_{{\rm A}i}\sigma_{{\rm B}j}
- J_{\rm A2}\sum_{<i,j>}\sigma_{{\rm A}i}\sigma_{{\rm A}j}  \nonumber
 \\
& & - J_{\rm B2}\sum_{<i,j>}\sigma_{{\rm B}i}\sigma_{{\rm B}j} 
 - H \sum_{i=1}^{N} \sigma_{{\rm A}i} -H \sum_{i=1}^{N} \sigma_{{\rm
 B}i} \nonumber \\
& & + Nz_6J_1 + Nz_3(J_{\rm A2}+J_{\rm B2})/2 , \nonumber \\ 
 H &=&  \epsilon_{\rm A}/2 
%\tag{2.7},
\eeqa
where $z_6=3$ and $z_3=6$ are the coordination numbers of honeycomb
 lattice and triangular lattice, respectively.

\section{Imaginary Path-Weight Method}
We calculate interface quantities by a random walk method with imaginary
 path-weight(IPW)\cite{holzer,akutsu90,akutsu90j}.
 We regard an interface with zigzag configuration as a trace of 2D
 {\it free} random walk.    

Consider an interface  which connects site $O$ and $P$.  
We denote the distance between site $O$ and $P$ by $R$ (Fig. \ref{fig0}).
The interface of the two-dimensional Ising model is made by fixing the boundary spins as depicted in Fig. \ref{fig0}.
Let us denote the partition function the Ising model with and without
 interface by $Z_R^{+-}(\theta)$ and  $Z_R^{++}$, respectively, 
where $\theta$ is the slant angle of an interface relative to a 
lattice axis. \cite{zia78,abraham86}
The interface tension $\gamma(\theta,T)$ is defined as
\beq
\gamma(\theta,T)= - \kBT \lim_{R \rightarrow \infty} \frac{1}{R}\ln 
\left [ \frac{Z_R^{+-}(\theta)}{Z_R^{++}}\right ],
\eeq
where $Z_R^{+-}(\theta)/Z_R^{++}$ is regarded as the interface partition 
function ${\cal G}$.

We apply Vdovichenko's method\cite{Vdo} to deal with the low-temperature 
diagrammatic expansion 
of $Z_R^{+-}(\theta)$ and  $Z_R^{++}$.
The method, which originally treated the high-temperature expansion of the 
partition function, also works for low-temperature expansion to evaluate 
weighted sum over all possible domain-wall configurations.  
The essential point in the Vdovichenko's method is introduction of the imaginary factor $e^{i \phi /2}$ at each turn (with angle $\phi$) of the random walks of the domain wall.
With this simple recipe, the problem reduces to a free random walk problem on a lattice.

We see that $Z_R^{++}$ equals weighted sum over possible configurations of 
closed domain walls; and $Z_R^{+-}(\theta)$  equals weighted sum over possible
 configurations of closed domain walls plus a single ``open'' domain wall 
traversing the lattice.
 In evaluating $Z_R^{+-}(\theta)$ by Vdovichenko's method, the free random walk
 nature allows us to ``decouple'' the open domain wall from closed domain 
walls.\cite{holzer,akutsu90}
Therefore, in the limit of $R \rightarrow \infty$, the interface partition 
function is equivalent to  the ``edge-to-edge'' lattice Green's function of
 the free random walk with running on the dual lattice.\cite{akutsu90}
Thus, in the limit of $R \rightarrow \infty$,  the interface partition
 function $\G$ is written as\cite{akutsu90}
\beqa
\G &=& \exp (- \gamma(\theta) R/\kBT)  \nonumber \\
&=&  \frac{1}{(2\pi )^2}\int_{-\pi}^{\pi}\int_{-\pi}^{\pi}{\rm d}k_{x}{\rm
 d}k_{y}
\frac{{\rm e}^{{\rm i}\svec{k}\svec{R}}}{D(\vec{k})} ,
%\tag{3.1}
\eeqa
where the $D$-function
  is defined as
\beq  
D(\vec{k})=\det[1-\hat{A}(\vec{k})]. 
\label{dfunction}
 %\tag{3.2}
\eeq
Here,  $\hat{A}(\vec{k})$ is the Fourier component of the connectivity
 matrix $A(\vec{r})$ which characterizes the random walk.

The above-described imaginary path-weight random walk method to calculate $\gamma(\theta,T)$ is exact only for solvable cases.  However, we have verified that the method works fairly well also for non-solvable cases\cite{akutsu97,akutsu90j,akutsu92,akutsu95}.

After evaluating the integral by the pure imaginary saddle point
  $\vec{\omega}=(\omega_x, \omega_y$),  we obtain a set of equations\
 as
\beqa
& & D({\rm i}\vec{\omega})  =   0,  \nonumber \\
%\tag{3.3a}\\ 
& & \frac{\del D({\rm i}\vec{\omega})}{\del \omega_{y}}/\frac{\del
 D({\rm i}\vec{\omega})}{\del \omega_{x}}  =  \tan\theta,
%\tag{3.3b}
\label{tantheta}
\eeqa
and
\beq
\gamma(\theta,T) =  \kBT(\omega_{x}\cos \theta+\omega_{y}\sin   \theta).
%\tag{3.4}
\label{gamma}
\eeq
From the thermodynamical theory on equilibrium crystal shape(island
 shape)\cite{andreev}, we have,
\beq
\omega_x = \lambda y/\kBT,  \qquad \omega_y = \lambda x/\kBT,
 %\tag{3.5}
\label{oxoy}
\eeq
where $\lambda$ is the Lagrange multiplier associated with the volume-fixing
 constraint in the Wulff construction, and $x$ and $y$ are the Cartesian
 coordinates describing the 2D island shape.  
Thus, we obtain the island shape directly from (\ref{tantheta}) with (\ref{gamma}).
  Eq. (\ref{oxoy}) gives relation between the interface orientation angle
 $\theta$ and the point $(x, y)$ on the  island shape.

The interface stiffness, which we denote by $\tilde{\gamma}(\theta)$,
 is given by\cite{akutsu92}
\beqa
\tilde{\gamma}(\theta) &=& \gamma(\theta) + \frac{\del^2
 \gamma(\theta)}{\del
 \theta^2}   \nonumber \\
&=& \kBT\sqrt{D_x^2 +D_y^2} \cdot
 [ -D_{xx}\sin   ^2 \theta \nonumber \\
& & +D_{xy}\sin    2 \theta 
-D_{yy}\cos ^2 \theta  ]^{-1}, 
%\tag{3.6a}
\label{stiffness}
\eeqa
where
\beqa
D_x=\frac{\del D}{\del \omega_x}, \quad D_y=\frac{\del D}{\del \omega_x},
 \nonumber \\[2mm]
D_{xx}= \frac{\del^2 D}{\del \omega_x^2}, \quad D_{yy}= \frac{\del^2
 D}{\del \omega_y^2}, \nonumber \\[2mm]
D_{xy}= n\frac{\del^2D}{\del \omega_x \omega_y}. 
%\tag{3.6b}
\label{stiffness1}
\eeqa\\

The one-dimensional interface of the lattice gas corresponds to a
 step on the vicinal surface.  For the vicinal surfaces, the surface
 free energy per projected area, which we denote by $f(\rho)$, is
 written as\cite{gmpt,akutsu88,yamamoto88}
\beqa
f(\rho)&=& f(0)+ \gamma(\theta) \rho +B(\theta) \rho^3,\nonumber
 \\
\rho&=& \frac{1}{a_h}\tan \phi  ,
%\tag{3.7}
\eeqa 
where $\rho$ is the step density, $\phi$ is the tilted angle of the
 vicinal surface, and $a_h$ is the step height, and $B(\theta)$ is
   the inter-step interaction coefficient. 
 We have\cite{yamamoto94,williams93}
 
\beqa
B(\theta)&=&\frac{\pi^2 }{6} \frac{(\kBT)^2}{\tilde{\gamma}(\theta)}
 \lambda^2(g_0),  \nonumber \\
\lambda(g_0)&=&\frac{1}{2}\left(1+ \sqrt{1+\frac{4
 \tilde{\gamma}(\theta)}{(\kBT)^2}
 g_0 }\right).
\label{BandA}
%\tag{3.8}
\eeqa
where $g_0$ is the coupling constant of the long range interaction
 between the steps of the form $g_0/r_s^2$ ($r_s$ is the step separation
 distance). 
Note that, in the limit of $g_0 \rightarrow 0$, the factor $\lambda
 (g) $ approaches unity, leading to\cite{akutsu88,yamamoto88}
\beq
B(\theta)=\frac{\pi^2 }{6} \frac{(\kBT)^2}{\tilde{\gamma}(\theta)}
 .
%\tag{3.9}
\label{B}
\eeq
Hence, the stiffness (\ref{stiffness}) can be utilized in determining the inter-step
 interaction coefficient $B(\theta)$.

\section{The Connectivity Matrix and  the $D-$function}
\subsection{The honeycomb lattice gas model with next nearest neighbor
 interaction}%
We apply the IPW method to calculate interface quantities of the
 nnn Ising model on the honeycomb lattice described by Hamiltonian
 (2.1,2.7) (Fig. \ref{fig1}). 
The Fourier components of the connectivity matrix $(A_{m,n})$ are
\beqa
A_{1,1}&=& \exp({\rm  i} k_x ) W/W_H  , \nonumber \\
A_{2,1}&=& \exp({\rm i} k_x ) W/W_H r_p   , \nonumber \\
A_{3,1}&=& \exp({\rm i} k_x ) W/W_H r_p  r_p  W_a  , \nonumber \\
A_{5,1}&=& \exp({\rm i} k_x ) W/W_H r_m  r_m  W_b  , \nonumber \\
A_{6,1}&=& \exp({\rm i} k_x ) W/W_H r_m    , \nonumber \\
A_{1,2}&=& \exp({\rm i} k_x /2+{\rm i} k_y  c_y  ) W\cdot W_H r_m
    , \nonumber \\
A_{2,2}&=& \exp({\rm i} k_x /2+{\rm i} k_y  c_y  ) W\cdot W_H   ,
 \nonumber \\
A_{3,2}&=& \exp({\rm i} k_x /2+{\rm i} k_y  c_y  ) W\cdot W_H r_p
    , \nonumber \\
A_{4,2}&=& \exp({\rm i} k_x /2+{\rm i} k_y  c_y  ) W\cdot W_H r_p
  r_p  W_b   , \nonumber \\
A_{6,2}&=& \exp({\rm i} k_x /2+{\rm i} k_y  c_y  ) W\cdot W_H r_m
  r_m  W_a   , \nonumber  \nonumber \\
A_{1,3}&=& \exp(-{\rm i} k_x /2+{\rm i} k_y  c_y  ) W/W_H r_m  r_m
  W_b   , \nonumber \\
A_{2,3}&=& \exp(-{\rm i} k_x /2+{\rm i} k_y  c_y  ) W/W_H r_m   
 , \nonumber \\
A_{3,3}&=& \exp(-{\rm i} k_x /2+{\rm i} k_y  c_y  ) W/W_H   , \nonumber
 \\
A_{4,3}&=& \exp(-{\rm i} k_x /2+{\rm i} k_y  c_y  ) W/W_H r_p   
 , \nonumber \\
A_{5,3}&=& \exp(-{\rm i} k_x /2+{\rm i} k_y  c_y  ) W/W_H r_p  r_p
  W_a   , \nonumber \\
A_{2,4}&=& \exp(-{\rm i} k_x ) W\cdot W_H r_m  r_m  W_a   , \nonumber
 \\
A_{3,4}&=& \exp(-{\rm i} k_x ) W\cdot W_H r_m    , \nonumber \\
A_{4,4}&=& \exp(-{\rm i} k_x ) W\cdot W_H   , \nonumber \\
A_{5,4}&=& \exp(-{\rm i} k_x ) W\cdot W_H r_p    , \nonumber \\
A_{6,4}&=& \exp(-{\rm i} k_x ) W\cdot W_H r_p  r_p  W_b   , \nonumber
 \\
A_{1,5}&=& \exp(-{\rm i} k_x /2-{\rm i} k_y  c_y  ) W/W_H r_p  r_p
  W_a   , \nonumber \\
A_{3,5}&=& \exp(-{\rm i} k_x /2-{\rm i} k_y  c_y  ) W/W_H r_m  r_m
  W_b   , \nonumber \\
A_{4,5}&=& \exp(-{\rm i} k_x /2-{\rm i} k_y  c_y  ) W/W_H r_m   
 , \nonumber \\
A_{5,5}&=& \exp(-{\rm i} k_x /2-{\rm i} k_y  c_y  ) W/W_H   , \nonumber
 \\
A_{6,5}&=& \exp(-{\rm i} k_x /2-{\rm i} k_y  c_y  ) W/W_H r_p   
 , \nonumber \\
A_{1,6}&=& \exp({\rm i} k_x /2-{\rm i} k_y  c_y  ) W\cdot W_H r_p
    , \nonumber \\
A_{2,6}&=& \exp({\rm i} k_x /2-{\rm i} k_y  c_y  ) W\cdot W_H r_p
  r_p  W_b   , \nonumber \\
A_{4,6}&=& \exp({\rm i} k_x /2-{\rm i} k_y  c_y  ) W\cdot W_H r_m
  r_m  W_a   , \nonumber \\
A_{5,6}&=& \exp({\rm i} k_x /2-{\rm i} k_y  c_y  ) W\cdot W_H r_m
    , \nonumber \\
A_{6,6}&=& \exp({\rm i} k_x /2-{\rm i} k_y  c_y  ) W\cdot W_H   ,
 \nonumber \\
%
%232
{\rm others} &=&0,
%\tag{4.1}
\eeqa
where   ${\rm i}^2=-1$, $c_y =\sqrt{3}/2$, $W=\exp [-2(J_1+2J_{\rm
 A2}+2J_{\rm B2})/(\kBT)]$, $W_H=\exp [-2H/(3\kBT)]$, 
$W_a=\exp [4J_{\rm A2}/(\kBT)]$, $W_b=\exp [4J_{\rm B2}/(\kBT)]$,
 $r_p=\exp({\rm i}\pi /6)$ and $r_m=\exp({\rm -i}\pi /6)$.

Then, the  $D$-function defined by (\ref{dfunction}) is
\beqa
D&&(k_x , k_y )=M + c_1 \cosh (k_x ) +    \nonumber \\
&&c_1 \cosh (k_x /2 - c_y  k_y ) +  
  c_1 \cosh (k_x /2 + c_y  k_y ) + \nonumber \\
&& c_2 \cosh (2 k_x ) +  
  c_2 \cosh (k_x  - 2 c_y  k_y ) +  \nonumber \\
&&c_2 \cosh (k_x  + 2 c_y  k_y ) +  
  c_3 \cosh (2 c_y  k_y ) +  \nonumber \\
&&c_3 \cosh (3 k_x /2 - c_y  k_y ) +  
  c_3 \cosh (3 k_x /2 + c_y  k_y ) +  \nonumber \\
&&  s_1 \sinh (k_x ) + s_2 \sinh (2 k_x ) -  \nonumber \\
&&s_2 \sinh (k_x  - 2 c_y  k_y ) +  
  s_4 \sinh (k_x /2 - c_y  k_y ) + \nonumber \\
&& s_4 \sinh (k_x /2 + c_y  k_y ) -  
  s_2 \sinh (k_x  + 2 c_y  k_y )  ,
%\tag{4.2}
\eeqa
where
\beqa
M&=&1 + 3 W^2 + 4 W^6 - W^3/W_H^3 - W^3 W_H^3 - \nonumber \\
&& 12 W^6 W_a +  
  9 W^6 W_a^2 + 4 W^6 W_a^3 +  \nonumber \\
&&(W^3 W_a^3)/W_H^3 +   
  W^3 W_H^3 W_a^3 - 6 W^6 W_a^4 + \nonumber \\
&& W^6 W_a^6 - 12 W^6 W_b +    30 W^6 W_a W_b +  \nonumber \\
&&(3 W^3 W_a W_b)/W_H^3 + 3 W^3 W_H^3 W_a W_b -  \nonumber \\
&&  18 W^6 W_a^2 W_b - 6 W^6 W_a^3 W_b + \nonumber \\
&& 6 W^6 W_a^4 W_b +    9 W^6 W_b^2 - \nonumber \\
&& 18 W^6 W_a W_b^2 + 3 W^4 W_a^2 W_b^2 +   \nonumber \\
&&  9 W^6 W_a^2 W_b^2 + 4 W^6 W_b^3 +\nonumber \\
&& (W^3 W_b^3)/W_H^3 +     W^3 W_H^3 W_b^3 - \nonumber \\
&& 6 W^6 W_a W_b^3 + 2 W^6 W_a^3 W_b^3 -   \nonumber \\
&&  6 W^6 W_b^4 + 6 W^6 W_a W_b^4 + W^6 W_b^6  ,
%\tag{4.3}
\eeqa
\beqa
c_1&=&W^2/W_H^2 - W^4/W_H^2 -   \nonumber \\
&& W/W_H - W W_H + W^2 W_H^2 -   \nonumber \\
&&    W^4 W_H^2 + (W^4 W_a)/W_H^2 + W^4 W_H^2 W_a +   \nonumber \\
&&     (W^4 W_a^2)/W_H^2 + W^4 W_H^2 W_a^2 - \nonumber \\
&&(W^4 W_a^3)/W_H^2 -  W^4 W_H^2 W_a^3 + \nonumber \\
&&(W^4 W_b)/W_H^2 + W^4 W_H^2 W_b -   \nonumber \\
&&     (W^2 W_a W_b)/W_H^2 - (W^4 W_a W_b)/W_H^2 +   \nonumber \\
&&     (2 W^3 W_a W_b)/W_H - (2 W^5 W_a W_b)/W_H +   \nonumber \\
&&     2 W^3 W_H W_a W_b - 2 W^5 W_H W_a W_b -  \nonumber \\
&&W^2 W_H^2 W_a W_b -   \nonumber \\
&&     W^4 W_H^2 W_a W_b + (3 W^5 W_a^2 W_b)/W_H +   \nonumber \\
&&     3 W^5 W_H W_a^2 W_b - (W^5 W_a^4 W_b)/W_H -   \nonumber \\
&&     W^5 W_H W_a^4 W_b + (W^4 W_b^2)/W_H^2 +  \nonumber \\
&&W^4 W_H^2 W_b^2 +    \nonumber \\
&&    (3 W^5 W_a W_b^2)/W_H + 3 W^5 W_H W_a W_b^2 -   \nonumber \\
&&     (3 W^5 W_a^2 W_b^2)/W_H - 3 W^5 W_H W_a^2 W_b^2 -   \nonumber
 \\
&&     (W^4 W_b^3)/W_H^2 - W^4 W_H^2 W_b^3 -  \nonumber \\
&&(W^5 W_a W_b^4)/W_H - W^5 W_H W_a W_b^4  ,
%\tag{4.4}
\eeqa
\beq
c_2=[W^3 (1 + W_H^2) (-1 + W_a) (-1 + W_b)]/W_H  ,
%\tag{4.5}
\eeq
\beqa
c_3&=& 
2 W^4 (-1 + W_a) (-1 + W_b) (-1 +  \nonumber \\
&&W_a + W_b + W_a W_b)  ,
%\tag{4.6}
\eeqa
\beqa
s_1&=&-[W (-1 + W_H) (1 + W_H) (W -    \nonumber \\
&& W^3 + W_H + W W_H^2 - W^3 W_H^2 +   \nonumber \\
&&         W^3 W_a + W^3 W_H^2 W_a + W^3 W_a^2 + \nonumber \\
&& W^3 W_H^2 W_a^2 -  W^3 W_a^3 - W^3 W_H^2 W_a^3 + \nonumber \\
&& W^3 W_b + W^3 W_H^2 W_b -   \nonumber \\
&&         W W_a W_b - W^3 W_a W_b - \nonumber \\
&& 2 W^2 W_H W_a W_b +     2 W^4 W_H W_a W_b - \nonumber \\
&&  W W_H^2 W_a W_b - W^3 W_H^2 W_a W_b -   \nonumber \\
&&         3 W^4 W_H W_a^2 W_b +W^4 W_H W_a^4 W_b + \nonumber \\
&& W^3 W_b^2 +   W^3 W_H^2 W_b^2 - \nonumber \\
&& 3 W^4 W_H W_a W_b^2 +  3 W^4 W_H W_a^2 W_b^2 -\nonumber \\
&&  W^3 W_b^3 -W^3 W_H^2 W_b^3 +  \nonumber \\ 
&&         W^4 W_H W_a W_b^4)]/W_H^2  ,
%\tag{4.7}
\eeqa
\beqa
s_2=-[W^3 (-1 + W_H) (1 + W_H) (-1  \nonumber \\
+ W_a) (-1 + W_b)]/W_H  ,
%\tag{4.8}
\eeqa
\beqa
s_4&=&[W (-1 + W_H) (1 + W_H) (W -  \nonumber \\
&& W^3 + W_H + W W_H^2 - W^3 W_H^2 +   \nonumber \\
&&       W^3 W_a + W^3 W_H^2 W_a +  \nonumber \\
&& W^3 W_a^2 + W^3 W_H^2 W_a^2 -   \nonumber \\
&&       W^3 W_a^3 - W^3 W_H^2 W_a^3 +  \nonumber \\
&& W^3 W_b + W^3 W_H^2 W_b -   \nonumber \\
&&       W W_a W_b - W^3 W_a W_b -  \nonumber \\
&& 2 W^2 W_H W_a W_b + 2 W^4 W_H W_a W_b -   \nonumber \\
&&       W W_H^2 W_a W_b - W^3 W_H^2 W_a W_b -  \nonumber \\
&& 3 W^4 W_H W_a^2 W_b +   W^4 W_H W_a^4 W_b +  \nonumber \\
&& W^3 W_b^2 + W^3 W_H^2 W_b^2 -   \nonumber \\
&&       3 W^4 W_H W_a W_b^2 + 3 W^4 W_H W_a^2 W_b^2 -  \nonumber
 \\
&& W^3 W_b^3 -  W^3 W_H^2 W_b^3 +  \nonumber \\
&& W^4 W_H W_a W_b^4)]/W_H^2  .
%\tag{4.9}
\eeqa

We substitute the $D$-function into (\ref{tantheta}) and solve them with respect
 to $(\omega_x,\omega_y)$ as a function of $\theta$.  Substituting
 the solution $(\omega_x(\theta),\omega_y(\theta))$ into (\ref{gamma})--(\ref{stiffness1}),
 we obtain the interface tension, 2D island shape  and the interface
 stiffness. 

Note that the $D$-function has the mirror symmetry with respect to
 $k_x$-axis, {\it i. e.} $D(k_x , k_y ) =  D(k_x , -k_y )$.
Therefore, the island shape  has the mirror symmetry with respect
 to $k_x$-axis. That is, $\omega_y(0)=0$ and $\omega_y(\pi)=0$ are
 the solutions of (\ref{tantheta}).  At the orientation corresponding to $\theta=0$
 or $\theta=\pi$, the form (4.2) reduces to
\beqa
D&&(k_x , 0)=M + c_3 + c_1 \cosh (k_x ) + \nonumber \\
&& 2 c_1 \cosh (k_x /2 ) +  
+ c_2 \cosh (2 k_x ) +  \nonumber \\
&&   2 c_2 \cosh (k_x  ) +  
   2 c_3 \cosh (3 k_x /2 ) +  \nonumber \\
&&   s_1 \sinh (k_x ) + s_2 \sinh (2 k_x ) - \nonumber \\
&&  2 s_2 \sinh (k_x ) + 
 2 s_4 \sinh (k_x /2 ) . 
%\tag{4.10}
\eeqa
From the solution of  $D(k_x , 0)=0$ ($\omega_y (0) = \omega_y (\pi)
 = 0$), we obtain  $\cosh(\omega_x (0)/2)$ and $\cosh(\omega_x (\pi)/2)$.
Then, from (\ref{gamma}), step tension becomes
\beqa
\gamma(0)=2 \kBT {\rm cosh}^{-1}(\omega_x (0)/2), \nonumber \\
\gamma(\pi)=2 \kBT {\rm cosh}^{-1}(\omega_x (\pi)/2).
%\tag{4.11}
\eeqa
In the $T \rightarrow 0$ limit, step tensions (step free energy per
 lattice constant) becomes,
\beqa
\gamma(0)&=&{\rm Min}[2 J_1+4J_2+\frac{2}{3}H, \nonumber \\
&& 2(2 J_1+4J_2-\frac{2}{3}H)], \nonumber \\
\gamma(\pi)&=&{\rm Min}[2 J_1+4J_2-\frac{2}{3}H,\nonumber \\
&& 2(2 J_1+4J_2+\frac{2}{3}H)],
%\tag{4.11}
\eeqa
where ${\rm Min}[a,b]$ denotes the smaller one in $\{a,b\}$.

From (\ref{stiffness}), the step stiffness becomes,
\beqa
\tilde{\gamma}(0)&=&
\kBT [|
c_1 \sinh ( \omega_x (0)/2)+ \nonumber \\
&& (c_1+2c_2)\sinh(\omega_x (0))+ \nonumber \\
&& 3c_3 \sinh(3 \omega_x (0)/2)+
2c_2 \sinh(2 \omega_x (0)) + \nonumber \\
&& s_4 \cosh( \omega_x (0)/2)+
(s_1+2 s_2) \cosh (\omega_x (0))+ \nonumber \\
&& 2 s_2 \cosh (2 \omega_x (0))
|]/ 
[2 c_y ^2(
2 c_3 +  \nonumber \\
&& c_1 \cosh ( \omega_x (0)/2) + 
4 c_2 \cosh(\omega_x (0))+ \nonumber \\
&& c_3 \cosh(3\omega_x (0)/2)+  
s_4 \sinh(\omega_x (0)/2)- \nonumber \\
&& 4s_2 \sinh(\omega_x (0)))]  ,  \\[2mm]
\tilde{\gamma}(\pi)&=&
\kBT [|c_1 \sinh ( \omega_x (\pi)/2)+ \nonumber \\
&& (c_1+2c_2)\sinh(\omega_x (\pi))+ \nonumber \\
&& 3c_3 \sinh(3 \omega_x (\pi)/2)+
2c_2 \sinh(2 \omega_x (\pi)) + \nonumber \\
&& s_4 \cosh( \omega_x (\pi)/2)+
(s_1+2 s_2) \cosh (\omega_x (\pi))+ \nonumber \\
&& 2 s_2 \cosh (2 \omega_x (\pi))
|]/[
2 c_y ^2(
2 c_3 + \nonumber \\
&&  c_1 \cosh ( \omega_x (\pi)/2) + 
4 c_2 \cosh(\omega_x (\pi))+ \nonumber \\
&& c_3 \cosh(3\omega_x (\pi)/2)+ 
s_4 \sinh(\omega_x (\pi)/2)- \nonumber \\
&& 4s_2 \sinh(\omega_x (\pi))
)]  .
%\tag{4.12}
\eeqa
In Fig. \ref{fig2}, we show an example of equilibrium island shape and a polar
 graph of step stiffness at $J_1=165$meV, $J_2/J_1=-0.1$, lattice
 constant$=3.84\mbox{\AA}$ and $H/J_1=0.31$. We also show the temperature
 dependence of step tension, step stiffness, the coefficient of step
 interaction (\ref{B}) where $g=B/a_h^3$, $a_h=3.14\mbox{\AA}$ and $g_0=0$.

In the absence of nnn interactions, $W_a$ and $W_b$ reduce to unity.
  The  $D$-function (4.2), then, becomes
\beqa
D&&(k_x , k_y)=M + c_1 \cosh (k_x ) + \nonumber \\
&& 2 c_1 \cosh (k_x /2- c_y  k_y  ) +  
 c_1 \cosh (k_x /2 + c_y  k_y )+  \nonumber \\
&& s_1 \sinh (k_x ) +   
 s_4 \sinh (k_x /2 - c_y  k_y ) + \nonumber \\
&& s_4 \sinh (k_x /2 + c_y  k_y ), 
%\tag{4.13} 
\eeqa
\beqa
M &=& 1+3W^2+3W^4+W^6+4W^3(W_H^3+ \nonumber \\
&& 1/W_H^3), \nonumber \\
c_1&=& - (1-W)^2 W (1+W)^2(1+W_H^2)/W_H, \nonumber \\
s_1&=&-s_4 \nonumber \\
&=&(1-W)^2 W (1+W)^2(1+W_H)(1- \nonumber \\
&& W_H)/W_H, \nonumber \\
c_2&=&c_3=s_2=0,
%\tag{4.14}
\eeqa 
which agrees with the $D$-function given in Ref.\cite{akutsu92}.

\subsection{The case of $H=0$}

At $H=0$, $s_1$, $s_2$ and $s_4$ in (4.2) become zero, since $W_H$
 reduces to unity.  The $D$-function, then, has the symmetry $D(k_x
 , k_y)=D(\pm k_x , \pm k_y)$.  The island shape has the mirror symmetry
 with respect to the $k_y$-axis too; $\omega_x(\pi/2)=\omega_x(3\pi/2)=0$
 becomes the solution of (\ref{tantheta}). 
The equation 
\beqa
D(0 , k_y )&=&M + c_1 +c_2 + 
2(c_1+c_3 ) \cosh ( c_y  k_y ) +   \nonumber \\
&&  (2 c_2+c_3) \cosh (2 c_y  k_y )  \nonumber \\
&& =0 ,
%\tag{4.15}
\eeqa
is solved, in terms of $\cosh(c_y \omega_y)$ ($k_{y}= {\rm i}\omega_{y}$),
 as
\beqa
&& \cosh(c_y \omega_y)= -\frac{c_1+c_3}{2(2 c_2 +c_3)} + \nonumber
 \\
&& \sqrt{\frac{(c_1+c_3)^2}{4(2c_2+c_3)^2}+\frac{-c_1+c_2+
c_3-M}{2(2c_2+c_3)}}
 \nonumber \\
&& \equiv z  .
%\tag{4.16}
\eeqa
By use of this solution, we obtain  step tension as
\beqa
\gamma(\pi/2)&=&\gamma(3\pi/2) \nonumber \\
&=&\frac{\kBT}{c_y} \cosh^{-1}(z),
%\tag{4.17}
\eeqa
and the step stiffness as
\beqa
&& \tilde{\gamma}(\pi/2)=\tilde{\gamma}(3\pi/2) \nonumber \\
&& =\frac{-4 \kBT c_y \sqrt{z^2-1} |c_1+c_3 +2(2 c_2+c_3)z|}{2c_1+4c_2+
(c_1+9c_3)z+8c_2z^2}.
%\tag{4.18}
\eeqa

In the case of  $J_{\rm A2}= J_{\rm B2}= 0$, the $D$-function reduces
 to that of the exact solution for the nn honeycomb lattice 
system\cite{zia}.
 
That is,
\beqa
D(k_x , k_y )&=&M + c_1 \cosh (k_x ) +  c_1 \cosh (k_x /2 - c_y 
 k_y ) +   \nonumber \\
&&  c_1 \cosh (k_x /2 + c_y  k_y ), \nonumber 
\eeqa
\beqa
 M&=&(1+W)^2(1-2W+6W^2- \nonumber \\
&&   2W^3+W^4), \nonumber \\
 c_1&=&-2(1-W)^2W(1+W)^2, \nonumber \\
 c_2&=&c_3=s_1=s_2=s_4=0.
%\tag{4.19}
\eeqa
At $\theta=0$, we obtain an explicit form of the solutions as
 \beqa
\omega_y(0)&=&0,  \nonumber \\
\cosh(\omega_x(0)/2)&=&\frac{1}{2} \sqrt{3-2M/c_1}-
\frac{1}{2}. 
%\tag{4.20}
\eeqa
Also, at $\theta=\pi/2$, we have
\beqa
\omega_x(\pi/2)&=&0,  \nonumber \\
\cosh(c_y \omega_y(\pi/2))&=&\frac{1}{2} (-M/c_1-1).
%\tag{4.21}
\eeqa
Hence, the interface tensions become
\beqa
\gamma(0)&=&2\kBT\ln (m_y), \nonumber \\
&& m_y=
-\frac{1}{2} + \frac{1}{
    2 }z_2 + 
   \frac{1}{2}\sqrt{(z_2-3)(z_2+1)}, \nonumber \\
&&z_2=\frac{(1+W)\sqrt{1-W+W^2}}{\sqrt{W}|1-W|} 
\eeqa
\beq
\gamma(\pi/2)=1/c_y  \cdot \kBT \ln [(-2 + 1/W + W)/2].
%\tag{4.22}
\eeq
Due to the Wulff's theorem, $\gamma(0)$ and  $\gamma(\pi/2)$ give
 the linear size of the island shape along $x$- and $y$-direction.
The explicit form of interface stiffness becomes
\beqa
\tilde{\gamma}(0)&=&
\frac{\kBT}{c_y ^2} \frac{z_2 \sqrt{-3-2z_2+z_2}}{2(z_2-1)} \\
\tilde{\gamma}(\pi/2)&=&4\kBT c_y  \nonumber \\
&&\times \frac{ (1 + W^2) |1 - 4 W + W^2|}{
  1 + 4W - 6W^2 + 4W^3 + W^4}.
%\tag{4.23}
\eeqa

\subsection{Triangular lattice}
In the limit of $J_1 \rightarrow 0$, the system becomes two independent
 triangular lattice gases.  In this case, the random walk treatment
 on the honeycomb lattice may not be good enough.  In fact, (4.2)
 in this limit gives  $D$-function slightly different from the known
 exact one on the triangular lattice.  Hence, we  need a separate
 study to treat this case.

The $D$-function of the triangular lattice becomes
\beqa
D_3(k_x ,k_y )&=&M - 
  c_1 \cosh(  k_y ) -  
  c_2 \cosh(\sqrt{3} k_x /2 -   k_y/2 ) - \nonumber \\
&&  c_3 \cosh(\sqrt{3} k_x /2 +   k_y/2 ),
%\tag{4.24}
\eeqa
where 
\beqa
M&=&1 + W_1^2 W_2^2 + W_1^2 W_3^2 + W_2^2 W_3^2, \nonumber \\
c_1&=&2 (1 - W_1) (1 + W_1) W_2 W_3 ,  \nonumber \\
c_2&=&2 W_1 W_2 (1 - W_3) (1 + W_3) , \nonumber \\
c_3&=&2 W_1 (1 - W_2) (1 + W_2) W_3 ,
%\tag{4.25}
\eeqa
$W_1=\exp [-2 J_1/(\kBT)]$, $W_2=\exp [-2 J_2/(\kBT)]$, and $W_3=\exp
 [-2 J_3/(\kBT)]$.  The equations agree with the known exact one\cite{zia}.
Note that the form of $D_3(k_x,k_y)$ (4.27) is essentially the same
 as  $D(k_x,k_y)$ of nn honeycomb lattice (4.20). Therefore, the
 island shape of triangular lattice obtained from (\ref{tantheta}) is the same
 as the one of honeycomb lattice.  The difference is the temperature
 dependence of coefficients. 
When $J_1=J_2=J_3=J$, i.e., $W_1=W_2=W_3=W$, $D$-function has a symmetry
 of $D_3(k_x ,k_y )=D_3(\pm k_x ,\pm k_y )$.   Hence, we have explicit
 forms of $\gamma$ and $\tilde{\gamma}$ for special orientations.
Therefore, we have
\beqa
 && \gamma_3(0)=\frac{2}{\sqrt{3}}\kBT \ln [\frac{1 - W^2}{2W^2}],
  \\
 && \gamma_3(\pi/2)=2\kBT  \times \nonumber \\
 && {\rm cosh}^{-1}\left(-\frac{1}{2} +\frac{1}{2}\sqrt{3+\frac{1
+3W^4}{W^2(1-W)(1+W)}}\right).
%\tag{4.26}
\eeqa
The step stiffness is written as
\beqa
\tilde{\gamma}_3(0)&=&\frac{2\sqrt{3} \kBT (1+W^2)|1-3W^2|}{
 1+6W^2-3W^4 }  \\
\tilde{\gamma}_3(\pi/2)&=&\frac{2\kBT}{3} \times  \nonumber \\
&& \frac{\sqrt{(1+3W^2)(1+3W^4-2Wz_1+2W^3z_1)}}{W(z_1-W)|1-W^2|},
 \nonumber \\
&& z_1=\sqrt{\frac{1+3W^2}{1-W^2}}.
%\tag{4.27}
\eeqa

\section{Application to Si(111) surface}
\subsection{The 7$\times$7 reconstructed surface}
At temperatures lower than the $7\times 7 \leftrightarrow 1 \times
 1$ transition temperature ($\sim $1130 K), Si(111) surface forms
 7$\times$7 reconstructed structure called DAS (dimer adatom stacking-fault)
 structure\cite{takayanagi}.
The unit cell of the DAS structure is divided into the faulted half
 (FH) and the unfaulted half (UH)\cite{neddermeyer}, each of which
 forms a triangular lattice.
It has been observed that the step structure is well described by
 the combination of the FH unit and UH unit\cite{tochihara}.
We consider, therefore, a pair of triangular sublattices, where the
 one  represents the FH lattice system, and the other  represents
 the UH one.
Consequently, the system becomes equivalent to a stoichiometrically
 binary lattice gas on a honeycomb lattice with nn interactions, 
where the inequivalent sites of the lattice gas model are 
coarse-grained representations of these two halves of the $7\times 7$ 
unit cell.
Therefore, we can use (4.15) and (4.16) to calculated step quantities.
In Fig. \ref{fig1}, we regard closed circles as FH units, and open circles
 as UH units.  We set lattice constant $=3.84\times 7$ \AA, and step
 height $= 3.14$\AA.  We introduce the step running direction angle
 $\theta$ so that a straight step with $\theta=0$ corresponds to
 $(11\bar{2})$ step (Fig. \ref{fig3}).

In spite of the extensive experimental studies on Si(111)7$\times$7
 structure, the values of kink energy and step tension have not been
 settled yet. 
As one trial, we adopt the result of
 Eaglesham {\it et al.}\cite{eaglesham}
 where step tension $\gamma$ at 700 $^\circ$C was obtained from the
 equilibrium crystal shape (ECS) of Si: $\gamma_{111}= 5.7 \times
 10^{-11}$ J/m for (111) surface and $\gamma_{100}= 1.0 \times 10^{-11}$
 J/m for (100) surface.  Due to the $2\times1$ reconstruction, this
 value of $\gamma_{100}$ corresponds to the mean value of the $S_A$-step
 tension and the $S_B$-step tension, and is consistent with the one
 calculated in our previous papers\cite{akutsu97,akutsu98}.

We choose kink energy so that the calculated mean value of step tension
 for $(2\bar{1}\bar{1})$ and $(\bar{2}11)$ at 700 $^\circ$C reproduces
 the above mentioned value $5.7 \times 10^{-11}$ J/m ($=36$ meV/\AA).
  In addition, the experimental observation of the island shape (and
 also the shape of spiral step) gives further information on the
 kink energy, due to the Wulff's theorem for 2D ECS.  Let $h_{\vec{n}}$
 be the distance between the center (= Wulff point) of the ECS (island
 shape, in our case) and the tangential line of the ECS at a position
 on the ECS curve, where $\vec{n}$ is the interface normal vector
 at the position. The Wulff's theorem states that the ratio 
$h_{\vec{n}}/\gamma_{\vec{n}}$
 ($\gamma_{\vec{n}}$: interface tension, or step tension in our case)
 is $\vec{n}$-independent, leading to a relation 
$\gamma_{\vec{n}}/\gamma_{\vec{n}'}=h_{\vec{n}}/h_{\vec{n}'}$
 for arbitrary $\vec{n}$ and $\vec{n}'$. From the photographs of
 the experimental observation\cite{yagi,ichimiya}, we have 
$h_{\bar{2}11}/h_{2\bar{1}\bar{1}}=
 1.2$.  This ratio gives the ratio between the step free energies
 corresponding to these directions.  At such low temperature where
 the observation was made, these 
step free energies are well approximated
 by the step formation energies $(2J+2H/3)$ and $(2J-2H/3)$ (see
 (4.12)), giving us 
$h_{\bar{2}11}/h_{2\bar{1}\bar{1}}= 1.2=(2J+2H/3)/(2J-2H/3)$
 which amounts to $H/J=0.31$.

We then set $J=0.475$eV and $H=0.15$eV.  The kink energy  becomes
 1.05eV$=2J+2H/3$ for $(2\bar{1}\bar{1})$ step and 0.85eV$=2J-2H/3$
 for $(\bar{2}11)$ step, which are smaller than but is in the same
 order of magnitude of the ones reported in
 Refs.\cite{wilhelmi,williams93}.
  
The difference in the on-site energy between the UH and FH, is then
 $E_{FH}-E_{UH}= 4H= 0.59 \mbox{eV}$.
In the first principles study of Meade and Vanderbilt\cite{meade},
 surface energies of the Si(111) surface for various structures are
 calculated:  For example,  1.24 eV/$1\times1$ for the $2\times2$-adatom
 structure, and 1.27 eV/$1\times1$ for the $2\times2$-adatom (faulted)
 structure.  From these values, we can estimate $E_{FH}-E_{UH}$ to
 be $(1.27-1.24)\times 24 \sim 0.7\mbox{eV}$ which is in reasonable
 agreement with our value 0.59eV. 

In Fig. \ref{fig4}, we show equilibrium island shape at 400 $^\circ$C and
 850 $^\circ$C, and the temperature dependence of step tension, step
 stiffness and step interaction coefficient $g=B/a_h^3$ (see (\ref{B})).
 The step tension is almost constant below 1130K, because the temperature
 is very low as compared with the lattice-gas melting temperature
 of the model ($\sim$ 8300K). On the other hand, the step stiffness
 strongly depends on temperature in the same region.  The step stiffness
 for $(2\bar{1}\bar{1})$ step increases rapidly as temperature decreases.
  While, the step stiffness of $(0\bar{1}1)$ step becomes smaller
 and smaller and converges to zero at zero temperature.

Note that a similar analysis can be made for $n \times n$ DAS structure.
  From the photographs of small island of $5 \times 5$ 
structure\cite{ichimiya},
 we find that the island shape  has a six-fold rotational symmetry
 in contrast to the case of the $7 \times 7$ structure which has
 a three-fold rotational symmetry.  Recall that the six-fold rotational
 symmetry appear only when $H=0$ (see \S 4.2).  Therefore, the energy
 difference between FH and UH units in the case of $5 \times 5$ DAS
 structure is very small if exists.  We stress here that, also for
 other structures, observation of the anisotropy of the equilibrium
 island shape will be useful in determining the energy difference
 between FH and UH units.

\subsection{The 1$\times$1 high temperature surface }

The high-temperature Si(111) surface at about 900 $^\circ$C has the
 structure of the 1$\times$1 surface together with disordered adatoms
 with concentration of 0.25 \cite{khomoto,latyshev91,yang94}.  Further,
 Khomoto and Ichimiya\cite{khomoto} reported that the number ratio
 of adatoms sitting on T$_4$ site and H$_3$ site is 4:1.  Although
 the adatoms are considered to be in a disordered phase, the broad
 $\sqrt{3}\times \sqrt{3}$  peaks appear
 in diffraction observations\cite{iwasaki,khomoto}
 which suggests existence of the short-range order corresponding
 to formation of the hard-hexagon units\cite{sakamoto}.

For the high temperature surface of Si(111), experimental measurement
 of the step tension and the step stiffness has been a subject of
 active study\cite{williams93,bartelt93,alfonso,metois,latyshev}.
  The experimental values are, however,  not settled yet.  We make,
 therefore, several trial calculations for possible cases.  In all
 the cases, we choose the microscopic coupling constants so that
 the calculated step stiffness at 900$^\circ$C reproduces the value
 presented by Bartelt {\em et al}.\cite{bartelt93}.

\subsubsection{Case 1: Adatom in disordered phase without short-range
 order}

We use the honeycomb lattice gas system of (4.2), with $J_1$ and
 $J_2$ being regarded as effective coupling constants.

In Fig. \ref{fig1} we regard the filled circles as atoms of the top layer
 and the open circles as those of the second layer (the lattice constant
 $=3.84\mbox{\AA}$, the step height $a_h=3.14\mbox{\AA}$). 

We introduce the step running direction angle $\theta$ so that a
 straight step with $\theta=0$ corresponds to $(\bar{1}\bar{1}2)$
 step.  The effect of dangling bonds normal to (111) plane are taken
 into account by setting $H/J_1=1$.

We calculate equilibrium island shape, step stiffness and step interaction
 coefficient, which we show in Fig. \ref{fig5}.  Here, assuming that $J_{2}$
 is small, we have  set $J_2/J_1=0.2$ and $J_1=60$meV.  
The kink energy becomes 176meV for $(2\bar{1}\bar{1})$ step and 256meV
 for $(\bar{2}11)$ step.
 As is seen in the island shape and the polar graph of step stiffness
 at 900$^\circ$C, there remains anisotropy in the step stiffness
 in spite of the circular island shape. 
 The most significant characteristic
 of this results is the asymmetry between the orientations
 $(2\bar{1}\bar{1})$
 and $(\bar{2}11)$.  In contrast to the $7\times7$ structure, the
 stiffness as a function of the step orientation takes its maximum
 at $(\bar{2}11)$.
%%%

\subsubsection{Case 2: The 
$\radical"270370{3} \times \radical"270370{3}$
 short-range ordered phase}

In the case of $\sqrt{3}\times \sqrt{3}$-ordered phase of adatoms,
 we calculate  step quantities by using the triangular lattice gas
 model (Fig. \ref{fig6}a).  As has been pointed out in Sec. 4, the system
 has six-fold rotational symmetry.  We set the lattice constant to
 be $3.84 \times \sqrt{3} \mbox{\AA}$, and the step height to be
 $3.14 \mbox{\AA}$. We introduce the step running direction angle
 $\theta$ so that a straight step with $\theta=0$ corresponds to
  $(\bar{1}\bar{1}2)$ step.  The step tension and the step stiffness
 are calculated exactly from (4.30)--(4.33).  The effective coupling
 constant is obtained to be $J=62$meV (kink energy $= 248$meV). We
 show the calculated results in Fig. \ref{fig7}. The step stiffness takes
 its maximum at the orientation $\{10\bar{1}\}$.

\subsubsection{Case 3: The $2\times 2$ short-range ordered phase}
Calculation of step quantities can be done in the same fashion as
 in the case 2.  We set the lattice constant as
 $3.84 \times 2 \mbox{\AA}$,
 and the step height as $3.14 \mbox{\AA}$ (Fig. \ref{fig6}b).  We introduce
 the step running direction angle $\theta$ so that a straight step
 with $\theta=0$ corresponds to  $(0\bar{1}1)$ step.  The step stiffness
 takes its maximum at $\{\bar{2}11\}$.  The effective coupling constant
 is obtained to be $J=67$meV (kink energy $= 268$meV).  We show the
 calculated results in Fig. \ref{fig8}.

\section{Summary}
 We have considered the honeycomb lattice Ising system in a staggered
 field with both nearest-neighbor (nn) and next-nearest-neighbor
 (nnn) interactions, to calculate interface tension, interface stiffness,
 island shape by the imaginary path-weight (IPW) method.

 We have applied the calculated results to Si(111)
 7$\times$7-reconstructed
 surfaces and the high-temperature Si(111) 1$\times$1 surface. We
 have made estimation on the microscopic coupling constants from
 existing experimental data, and have drawn equilibrium island shape,
 step tension, step stiffness and the coefficient of step interaction,
 with their temperature dependence. Our analysis made in the present
 paper will be helpful in determining precise value of the kink energy
 from experimental observation.

Our lattice-gas treatment made in the present paper corresponds to
 the two-level approximation for the surface fluctuation.  Fortunately,
 the temperature-range  of our concern in the present study  is very
 low, the two-level approximation is expected to be fairly  reliable.
  On the other hand, at higher temperatures, near the roughening
 transition temperature, we should consider multilevel fluctuation
 of the surface.  Even in such cases, we have an efficient method,
 namely, the {\em temperature-rescaled 
Ising-model approach},\cite{akutsu98}
 where the IPW method is combined with the
 numerical renormalization-group
 method;\cite{NRG} details will be discussed elsewhere.

\acknowledgments
%{\large \bf Acknowledgement}\\

The authors thank  Prof. T. Yasue, Prof. E. D. Williams, Prof. A.
 Ichimiya, Dr. T. Suzuki and Prof. H. Iwasaki for helpful discussions.
 One of authors (N. A.) thanks Prof. T. Yasue for bibliographical
 information. The authors also thank  Prof.  
T. Nishinaga for encouragement.
 This work was partially supported by  the ``Research for the Future''
 Program  from The Japan Society for the Promotion of 
Science (JSPS-RFTF97P00201)
 and by the Grant-in-Aid for Scientific Research from Ministry of
 Education, Science, Sports and Culture (No.09640462).
\\

\begin{figure}
%\vspace*{5cm}
\caption{
Examples of an interface of square lattice  Ising model made by fixing the boundary spins.
}
\label{fig0}
\end{figure}

\begin{figure}
%\vspace*{5cm}
\caption{
Examples of an interface configuration. A-atom and B-atom are indicated
 by filled circle and open circle, respectively. Thick line represents
 an interface.
}
\label{fig1}
\end{figure}

\begin{figure}
%\vspace*{18cm}
\caption{
An example of calculation by the use of the $D$-function of (4.2).
 (a) The island shape at 900$^\circ$C, (b) a polar graph of step
 stiffness at 900$^\circ$C, (c) temperature dependence of step tension,
 (d) temperature dependence of step stiffness, and (e) temperature
 dependence of $g=B/a_h^3$.  We have set $J_1=165$meV, $J_2=-16.5$mev,
 and $H=165$meV: Kink energies are 88meV for $(2\bar{1}\bar{1})$
 step and 176meV for $(\bar{2}11)$ step.  In (c)-(e), thick lines
 correspond to$(\bar{2}11)$  step,  thin lines to $(2\bar{1}\bar{1})$
 step and broken lines to $\{ 10\bar{1} \}$ step.
}
\label{fig2}
\end{figure}

\begin{figure}[htbp]
%\vspace*{5.0cm}
\caption{
Examples of a step edge on $7\times 7$ reconstructed Si(111) surface.
 ``U'' denotes unfaulted half unit, and ``F'' denotes faulted half
 unit.
}
\label{fig3}
\end{figure}

\begin{figure}
%\vspace*{18cm}
\caption{
Calculation for $7 \times 7$ reconstructed surface by the use of
 the $D$-function of (4.15). (a) The island shape at 850$^\circ$C,
 (b) the island shape at 400$^\circ$C, (c) temperature dependence
 of step tension, (d) temperature dependence of step stiffness and
 (e) temperature dependence of $g=B/a_h^3$.  We have set $J=0.475$eV,
 $4H=0.59$eV: Kink energy $=1.05$eV for $(2\bar{1}\bar{1})$ step
 and 0.85eV for $(\bar{2}11)$ step.  In (c)-(e), thick lines corresponds
 to $(2\bar{1}\bar{1})$ step,  thin lines to $(\bar{2}11)$step and
 broken lines to $\{ 10\bar{1} \}$ step.
}
\label{fig4}
\end{figure}

\begin{figure}
%\vspace*{18cm}
\caption{
Calculation for $1 \times 1$  surface (Case 1) by the use of $D$-function
 of (4.2). (a) The island shape at 900$^\circ$C, (b) a polar graph
 of step stiffness at 900$^\circ$C, (c) temperature dependence of
 step tension, (d) temperature dependence of step stiffness and (e)
 temperature dependence of $g=B/a_h^3$ .
 We have set $J_1=60$meV,$J_2=12$mev,
 and $H=60$meV: Kink energy $=$ 176meV for $(2\bar{1}\bar{1})$ step
 and 256meV for $(\bar{2}11)$ step. In (c)-(e), thick lines corresponds
 to $(2\bar{1}\bar{1})$ step,  thin lines to $(\bar{2}11)$step and
 broken lines to $\{ 10\bar{1} \}$ step.
Open squares: Ref. [37].  Open circle: Ref. [35].
}
\label{fig5}
\end{figure}

\begin{figure}
%\vspace*{5cm}
\caption{
Examples of an interface configuration for two cases of adatom orderings:
 (a) $\radical"270370{3} \times \radical"270370{3}$ and (b) $2\times
 2$.  Thick line represents the interface.  A-atom and B-atom are
 indicated by filled circle and open circle, respectively.  Adatoms
 are indicated by shaded large circles.  Broken lines denote boundaries
 between the hexagons.
}
\label{fig6}
\end{figure}

\begin{figure}
%\vspace*{18cm}
\caption{
Case 2.  Calculation for $\radical"270370{3} \times \radical"270370{3}$
 adatom ordering (Case 2) by the use of $D$-function of (4.30)--(4.33).
  (a) The island shape at 900$^\circ$C, (b) a polar graph of step
 stiffness at 900$^\circ$C, (c) temperature dependence of step tension,
 (d) temperature dependence of step stiffness, (e) temperature dependence
 of $g=B/a_h^3$.  We have set $J=62$meV: Kink energy $=248$meV for
 $\{ 10\bar{1}\} $ step.  In (c)-(e), thick lines corresponds to
 $\{2\bar{1}\bar{1}\}$ step and thin lines to $\{ 10\bar{1} \}$ step.
Open squares: Ref. [37].  The open circle: Ref. [35].
}
\label{fig7}
\end{figure}

\begin{figure}
%\vspace*{18cm}
\caption{
Case 3.  Calculation for $2 \times 2$ adatom ordering (Case 3) by
 the use of $D$-function of (4.30)--(4.33).  (a) The island shape
 at 900$^\circ$C, (b) a polar graph of step stiffness at 900$^\circ$C,
 (c) temperature dependence of step tension, (d) temperature dependence
 of step stiffness and (e) temperature dependence of $g=B/a_h^3$.
  We have set $J=67$meV: Kink energy $=268$meV for ${2\bar{1}\bar{1}}$
 step.  In (c)-(e), thick lines correspond to $\{2\bar{1}\bar{1}\}$
 step and thin lines to $\{ 10\bar{1} \}$ step.
Open squares: Ref. [37].  The open circle: Ref. [35].
}
\label{fig8}
\end{figure}

\end{document}